\documentclass{article}

\usepackage{PRIMEarxiv}

\usepackage[utf8]{inputenc} 
\usepackage[T1]{fontenc}    
\usepackage{hyperref}       
\usepackage{url}            
\usepackage{booktabs}       
\usepackage{amsfonts}       
\usepackage{nicefrac}       
\usepackage{microtype}      
\usepackage{lipsum}
\usepackage{fancyhdr}       
\usepackage{graphicx}       
\graphicspath{{media/}}     

\title{Characterization of an aerosolized nanoparticle beam beyond the diffraction limit through strong field ionization}

\author{
  Michael Davino, Tobias Saule, Nora G. Helming, J.A. Powell, Carlos Trallero-Herrero* \\
  Department of Physics \\
  University of Connecticut \\
  Storrs, CT 06269\\
  \texttt{*carlos.trallero@uconn.edu} \\
}

\begin{document}
\maketitle

\begin{abstract}
The study of nanomaterials is an active area of research for technological applications as well as fundamental science. A common method for studying properties of isolated nanoparticles is by an in-vacuum particle beam produced via an aerodynamic lens. Despite being common practice, characterization of such beams has proven difficult as light scattering detection techniques fail for particles with sizes beyond the diffraction limit. Here we present a new technique for characterizing such nanoparticle beams using strong field ionization. By focusing an ultrafast, mJ-level laser into the particle beam, a nanoparticle within the laser focus is ionized and easily detected by its ejected electrons. This method gives direct access to the nanoparticle density at the location of the focus and, by scanning the focus through the transverse and longitudinal profiles of the particle beam, the 3-dimensional particle density distribution can be attained. Further, we show that strong field ionization is effective in detecting spherical nanoparticles as small as 10 nm in diameter. Additionally, this technique is an effective tool in optimizing the particle beam for specific applications. As an example we show that the particle beam density and width can be manipulated by restricting the gas flow into the aerodynamic lens.
\end{abstract}

\section{Introduction}

Nanoparticles (NPs) and their associated technologies are active areas of research across multiple scientific disciplines with important applications in medicine \cite{bobo2016nanoparticle,steichen2013review,rajendran2018review}, chemical analysis \cite{boken2017plasmonic,krajczewski2017plasmonic}, and solar energy \cite{nakayama2008plasmonic,kameya2011enhancement}. Apart from the variety of specific applications, the optical properties of NPs are an area of interest for fundamental scientific research \cite{boken2017plasmonic,ciappina2017attosecond,zhang2017nonlinear}. While much of the research and applications involve NPs embedded in substrates \cite{kravets2018plasmonic} or in dense aqueous suspensions \cite{kameya2011enhancement}, there is a need to study isolated NPs \cite{ringe2013single}. In the case of studying isolated aerosolized NPs (as compared to isolated NPs in a suspension) producing a target sample is particularly difficult. One common method for generating such a sample is by aerosolization of a NP suspension and subsequent collimation of the aerosol into an in-vacuum particle beam by means of an aerodynamic lens \cite{liu1995generating2}. This NP beam serves as the target for study.
\\
Though NP beams as produced by an aerodynamic lens have been thoroughly investigated numerically \cite{liu1995generating1,zhang2004numerical2,wang2005aerodynamic2}, experimental characterization has proven difficult as the various characterization techniques come with a range of limitations. In particular, light scattering techniques fail for beams comprised of NPs beyond the diffraction limit of the light source. Early characterizations were performed by detecting NP ions via a Faraday cup electrometer and using a knife edge to physically impede the NP beam, thus giving access to the transverse spatial distribution of the NPs \cite{liu1995generating2,wang2006experimental}. While this method can also measure the particle velocities by applying a transverse electric field to deflect the ionized NP beam, it has the obvious drawback of requiring the NPs to be ionized. More recently light-sheet imaging (LSI) has been used to directly image NPs in a beam by light scattered from a CW laser source \cite{worbs2019light}. This technique measures both the spatial distribution of the NPs and the exact number of particles per exposure time, all in a non-disruptive manner. However, being a scattering based technique, LSI has the significant drawback that it is limited to larger NPs (> 100 nm in \cite{worbs2019light}) due to diffraction limitations. Finally, laser-induced breakdown detection (LIBD) has also been utilized to characterize NP beams \cite{barreda2015situ}. The LIBD technique employs a pulsed laser focused into the NP beam and detects the presence of a NP in the focus by light emitted from its plasma. This detection scheme can be used to spatially resolve the number of NPs in a beam comparable to the capabilities of LSI but with the demonstrated capacity to detect particles as small as 5 nm \cite{lead2006environmental}.
\\
Here we present a new technique for acquiring spatially resolved NP density information for an aerosolized NP beam. By focusing an ultrafast mJ-level pulsed laser into a NP beam electrons are ejected via strong field ionization (SFI) and are easily detectable by a channel electron multiplier (CEM). Distinguishing laser pulses with a NP present in the focus gives direct access to the NP density of the beam at that position. Further, by scanning the laser focus through the transverse profile of the NP beam at multiple points along its axis of propagation, we attain the 3-dimensional NP density distribution of the particle beam. We show this scheme to be effective for spherical NPs of various materials and sizes, with NP diameters ranging from 10 nm – 120 nm.
\\
SFI is shown here to be a useful tool in monitoring the output of an aerodynamic lens setup, which is especially valuable in ensuring optimal and consistent experimental conditions. As an example, we use the technique to demonstrate that a NP beam may be manipulated by adjusting the flow of aerosol input to the aerodynamic lens, effectively allowing for the NP beam to be tailored to specific applications.

\section{Experimental Setup}

\begin{figure}[ht]
\centering
\includegraphics[width=\linewidth]{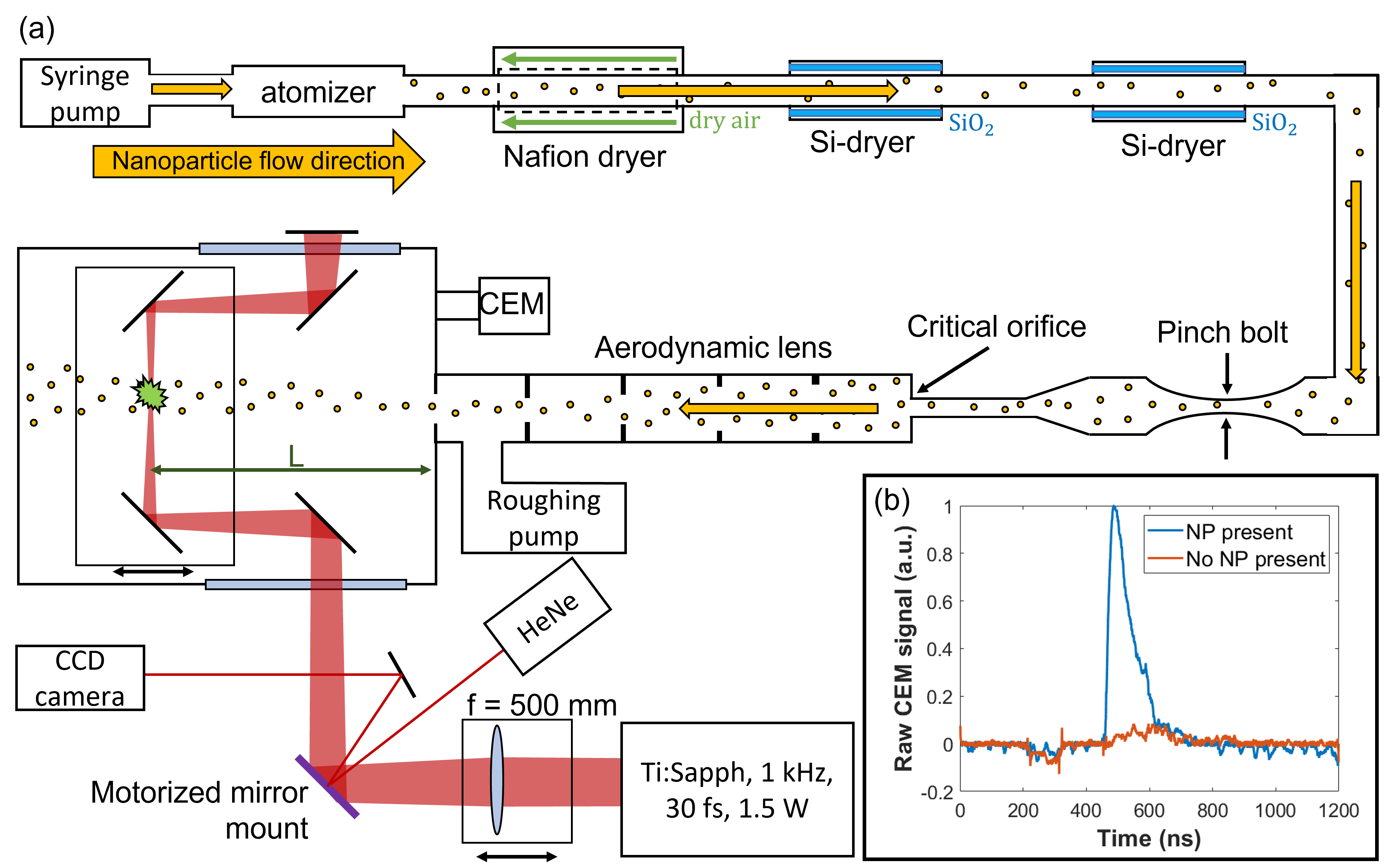}
\caption{(a) depicts the experimental setup. An atomizer is fed an aqueous NP suspension by a constant output syringe pump and the solution is aerosolized by nitrogen gas at a fixed pressure. The aerosol then passes through a series of dryers consisting of a Nafion™ membrane dryer and two silica bead dryers. The tubing that couples the output of the dryers to the aerodynamic lens is equipped with a pinch bolt that allows for choking the flow of gas and nanoparticles into the lens. The aerodynamic lens is kept at rough vacuum and focuses the dry nanoparticles into a beam which is sent into the interaction chamber which is equipped with an ion gauge and CEM. Outside of vacuum the output of a Titanium:Sapphire laser system is incident on a focusing lens mounted on a translational stage. The focusing beam reflects off a silver mirror mounted in a motorized mount and is sent into the interaction chamber. In vacuum the beam is steered by a series of dielectric mirrors mounted on a translational stage such that it focuses onto the NP beam a variable distance L from from the output of the aerodynamic lens. The laser is focused onto the nanoparticle beam a distance L from the exit of the aerodynamic lens. A HeNe laser is reflected off the motorized mirror and imaged with a CCD camera to track the position of the mirror. (b) shows a plot of the raw CEM signal for two laser pulses, one with a NP present in the interaction region and one without.}
\label{fig:setup}
\end{figure}

A schematic of the experimental setup is depicted in Figure \ref{fig:setup} (a). A well-established procedure for producing a collimated NP beam is utilized: an aqueous NP suspension is aerosolized by a syringe pump fed atomizer, dried to remove excess water, and injected into an aerodynamic lens where a series of apertures collimates the dry NPs into a beam. The setup used here is similar to the setup used by Powell et al.\cite{powell2019interplay}.

Two atomizers were employed in this work; the first is a design based on a TSI Inc. aerosol generator (model 3076) which produces an aerosol with an average droplet size of 0.3 $\mu$m. This atomizer uses impaction to remove large droplets from the final aerosol and thus a particularly fine mist is attained, however this means that much of the aerosol recondenses. For stable NP suspensions (e.g. SiO$_2$) this suspension can be collected and recycled back through the atomization process, but for less stable NPs the impaction process can be a catalyst for the NPs to aggregate. In general, we have found that metallic NPs are particularly susceptible to such aggregation and thus, for metallic samples a commercial, needle based, atomizer from Fuso Seiki Co. (Model ST-10SK-0.5X-G2) is used. This atomizer does not use an impactor to discriminate against large aerosol droplets, thus the average droplet size is larger, on the order of tens of microns, but allows for the use of the entire liquid sample without concerns of aggregation. Both atomizers were backed with 30 psi of nitrogen gas. The flow rate of liquid sample into the atomizers is kept constant by a syringe pump and set to 1 mL/min for the TSI atomizer and 0.1 mL/min for the Fuso atomizer.
\\
The output of either atomizer is coupled directly into a series of dryers used to remove the solvent (water) from the aerosol and leaving only the NPs suspended in gas to enter the aerodynamic lens. The first drying stage is a cylindrical Nafion™ membrane (5 inches diameter, 18 inches length) which operates by proton transport across the membrane assisted by a counterflow dry gas (compressed air, < 5\% relative humidity). This membrane dryer is followed by two silica bead diffusion dryers. The relative humidity of the suspended NPs is measured following the three dryers and is typically < 5\% for the TSI atomizer and < 15\% for the Fuso atomizer.
\\
The output of the dryers is coupled into the aerodynamic lens by an orifice (denoted ‘critical orifice’ in Figure \ref{fig:setup}) that has an inner diameter of 330 $\mu$m. The flow of NPs through the critical orifice can be limited by a pinch bolt installed on the hosing between the silica dryers and the orifice.

\begin{table}[ht]
\centering
\begin{tabular}{|l|l|l|l|l|}
\hline
Material & Size (nm) & Mass (g) & Aqueous Concentration (NP/mL) & Manufacturer \\
\hline
SiO$_2$ & 60 & 3.0e-16 & 3.3e14 & General Research and Engineering \\
\hline
SiO$_2$ & 120 & 2.4e-15 & 4.2e13 & General Research and Engineering \\
\hline
Ag & 10 & 5.5e-18 & 4.0e12 & Cytodiagnostics \\
\hline
Au & 120 & 1.8e-14 & 2.1e09 & Cytodiagnostics \\
\hline
ZnO & 30 & 7.9e-17 & 2.8e15 & Nanografi \\
\hline
TiO$_2$ & 33 & 8.0e-17 & 4.4e15 & Nanografi \\
\hline
Fe$_3$O$_4$ & 16 & 1.1e-17 & 2.1e16 & Nanografi \\
\hline
\end{tabular}
\caption{\label{tab:particles}Detailed NP information for the samples used. The concentration presented is the commercially available concentration, some samples were further diluted and are noted as such in the text.}
\end{table}

The aerodynamic lens consists of a series of six apertures of decreasing diameter. The pressure gradient across the lens is established by 1 atm input into the critical orifice and between 100 and 500 mTorr maintained at the exit of the lens by an Edwards Vacuum Drystar roughing pump and booster combination (Models QDP80 and QMB500F respectively). The NPs are collimated as they move through the aerodynamic lens and the output beam is sent through a 2 mm diameter skimmer into the interaction chamber which is kept between 8e-5 and 3e-3 Torr. The pressure in the interaction chamber is entirely dependent on the setting of the pinch bolt, and thus the ion gauge pressure sensor in the interaction chamber serves as a relative measure of the gas flow (and subsequently, NPs) into the aerodynamic lens. At a set position of the pinch bolt the chamber pressure is stable to ± 5e-6 Torr.
\\
The interaction chamber is equipped with a continuous dynode channel electron multiplier (CEM) set to detect electrons. The output CEM signal is put through a boxcar integrator with a gate width of 15 $\mu$s and subsequently recorded by a National Instruments measurement box which is controlled via LabVIEW.
\\
The laser used here is a Titanium:Sapphire system producing 1.5 mJ pulses at a rate of 1 kHz (1.5 W) with a duration of 30 fs at a central wavelength of 790 nm. The beam is sent through a 500 mm focusing lens mounted on a translational stage and subsequently reflected by a silver mirror mounted on a 1D motorized mount which injects the beam into the interaction chamber. In vacuum the beam is steered by a series of dielectric mirrors mounted on a translational stage such that it focuses onto the NP beam a variable distance L from from the output of the aerodynamic lens. The focal spot size was measured to be 45.8 ± 0.5 $\mu$m 1/e$^2$ diameter (peak intensity  on the order of 1e15 W/cm$^2$).
\\
The motorized mirror is used to scan the laser focus through the transverse profile of the NP beam. A Helium-Neon (HeNe) laser is utilized as a reference to track the position of the motorized mirror by detecting its reflection off the mirror with a CCD camera. By measuring the distances from the motorized mirror to both the focus and the camera, the relative displacement of the focus with respect to the NP beam may be attained.
\\
The NP samples used in this study are commercially available products purchased as an aqueous suspension and can be diluted to control the NP concentration. Complete details regarding the NP samples can be found in Table \ref{tab:particles}.

\section{Results}

\begin{figure}[!ht]
\centering
\includegraphics[width=\linewidth]{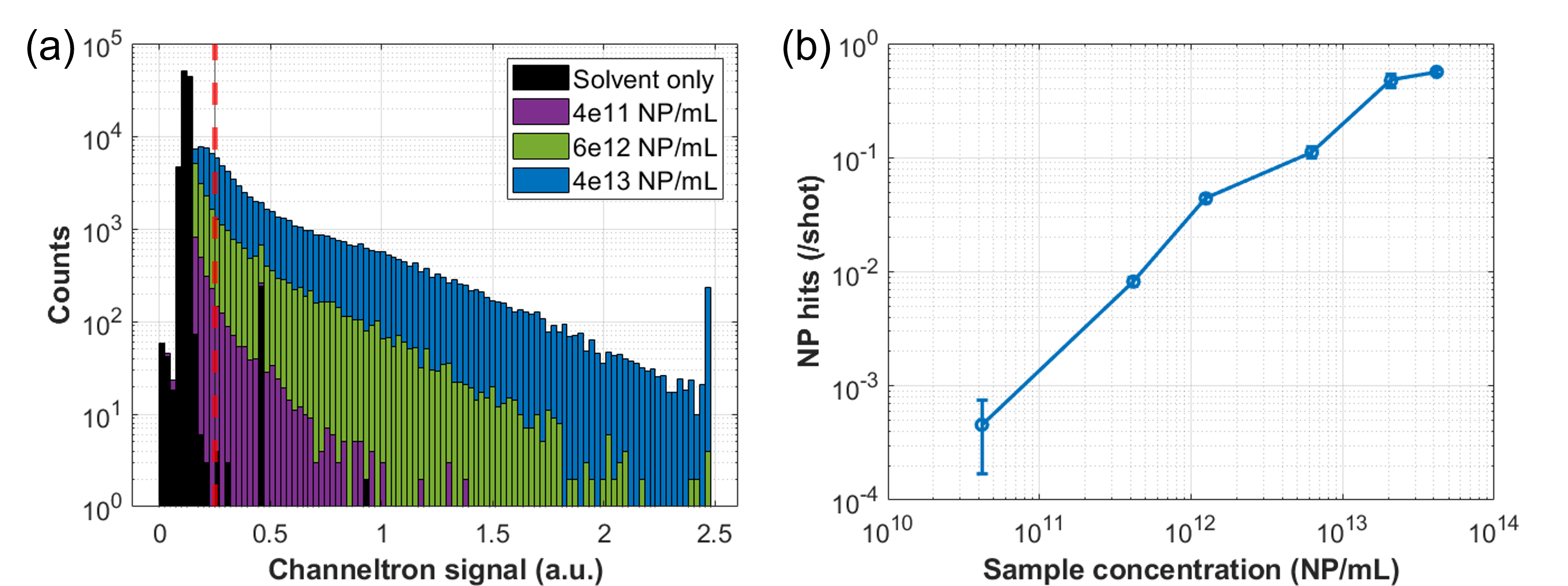}
\caption{Integrated CEM signal for various suspension concentrations of 120 nm SiO$_2$. (a) is a histogram of 1e5 laser shots for three NP concentrations as well as a solvent (water) only sample (i.e. 0 NP/mL). The dashed vertical red line indicates the threshold chosen to distinguish the presence or absence of a NP in the interaction region, CEM signals larger than the threshold are considered a NP hit. (b) plots the number of NP hits, normalized per laser shot, as a function of atomized (aqueous) NP sample concentration. Note the artifact present in the solvent only data set at a signal value of $\sim$0.5 in (a) is known to be due to transient noise in the CEM signal line. Error bars represent the standard deviation within the data set which becomes negligible at higher concentrations.}
\label{fig:concentration}
\end{figure}

A critical prerequisite in characterizing a NP beam by means of SFI is the ability to identify laser shots for which a NP is present in the interaction region. Inset Figure \ref{fig:setup} (b) shows the raw CEM signal for both the presence and absence of a NP in the interaction region. There is a clear distinction seen, but it is important to realize that not every NP ‘hit’ will necessarily generate as strong of a CEM signal. The number of ejected electrons has been shown to be strongly dependent on the peak intensity of the laser pulse impingent on the NP\cite{powell2019interplay}. Further, because the NP is several orders of magnitude smaller than the focal spot size and thus the local peak intensity will vary with position of the NP within the focal region\cite{powell2016intensity}. This means that the integrated CEM signals that are measured here (now referred to as simply the CEM signal) will constitute a continuous spectrum meaning that the distinction between the presence and absence of a NP in the focal region is non-trivial.
\\
To better identify the presence of a NP in the interaction region we have performed concentration dependent NP studies which are presented in Figure \ref{fig:concentration}. The laser focus was positioned at L = 5.5 cm and aligned to maximize the number of clearly distinguishable NP hits in order to focus (approximately) in the center of the NP beam. Figure \ref{fig:concentration} (a) is a histogram of the CEM signal for 1e5 laser shots focused into a NP beam consisting of 120 nm SiO$_2$ at three NP suspension concentrations spanning from 4e11 to 4e13 NP/mL as well as a solvent (water) only sample. Note that the artifact present in the solvent only data at a signal value of $\sim$0.5 in Figure \ref{fig:concentration} (a) is known to be due to transient noise in the CEM signal line. Because of this, subsequent data presented in this manuscript rejects laser pulses that records a CEM signal of this value.

\begin{figure}[t!]
\centering
\includegraphics[width=130mm]{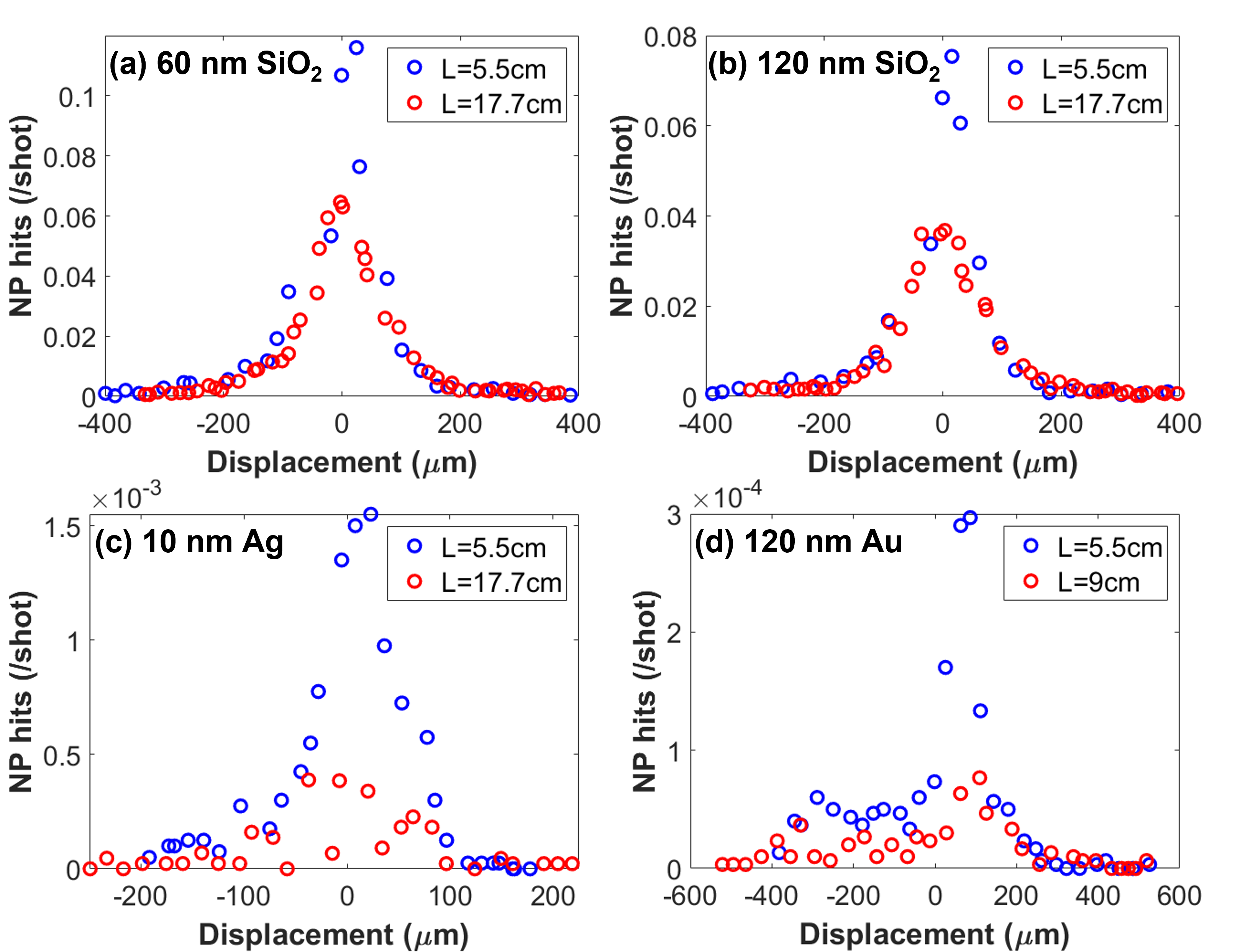}
\caption{1D scans of NP hits per laser shot for (a) 60 nm SiO$_2$, (b) 120 nm SiO$_2$, (c) 10 nm Ag, and (d) 120 nm Au taken with a main chamber pressure of 3.5e-4 Torr. Each subfigure presents scans for both the smallest and largest L value that a scan was performed at. The atomized NP concentration for (a) — (d) are, respectively: 5.0e13, 6.3e12, 4.0e12, and 2.1e9 NP/mL. CEM signals were recorded for 5e3, 5e3, 5e4, and 3e5 laser shots for each step of the motorized mirror in (a) — (d) respectively.}
\label{fig:profile}
\end{figure}

From Figure \ref{fig:concentration} (a) it is immediately evident that, when NPs are atomized, larger CEM signals are observed. As this is the case, we then use the CEM signal distribution attained from the solvent only and define a threshold that will be used to identify a NP hit. This threshold is chosen as a CEM signal value larger than those seen for the solvent only sample and is denoted by a dashed vertical red line in Figure \ref{fig:concentration} (a). We then say that any laser shot which produces a CEM signal larger than this threshold constitutes the presence of a NP in the interaction region. Figure \ref{fig:concentration} (b) shows the number of NP hits as a function of NP concentration produced by utilizing this analysis technique. The observed relation fits with the expectation that higher concentrations of the aqueous suspension directly correlates to a higher in-vacuum hit rate.
\\
We must now note some of the limitations associated with this NP detection scheme. Examination of Figure \ref{fig:concentration} (a) shows that, with NPs being atomized, there are a significant amount of laser shots for which the CEM signal is larger than the average signal from water only, but smaller than the threshold value. This is to say that there are a significant number of laser shots where there is likely a NP present in the interaction region (or more likely, on its edges) which are not being counted as such. This means that the number of NPs counted by this scheme represents a lower bound to the actual NP count. This is compounded with the realization that this scheme, as described, does not differentiate the presence of a single NP or multiple NPs within the interaction region for a single laser shot. Instead, we are counting a maximum of a single NP per laser shot when there may be more. For these reasons the number of NPs detected in-vacuum will be an undercount to the actual number.
\\
With these limitations in mind, we now aim to quantify the NP beam width. This is accomplished by scanning the laser focus through the transverse profile of the NP beam using the motorized mirror which injects the laser into the interaction chamber (see Figure \ref{fig:setup}). The reference HeNe allows for a change in angle of the motorized mirror to be measured and subsequently correlated to a relative displacement of the laser focus. By measuring the number of NP hits per laser shot at many points along the transverse profile of the NP beam, the relative density of the beam is effectively mapped in one dimension. Figure \ref{fig:profile} shows a series of such scans for samples of SiO$_2$, Ag, and Au. It is immediately evident that at L = 5.5 cm each of the four samples produced a NP beam, each with a width on the order of hundreds of microns. Further, the beam width is largely preserved through L = 17.7 cm (L = 9 cm for Figure \ref{fig:profile} (d)), but the total number of NPs hit decreases as the laser focus moves further from the exit of the aerodynamic lens. We also note that Figure \ref{fig:profile} (a) and (b) show that both 60 nm and 120 nm SiO$_2$ retain high-quality beam profiles to L = 17.7 cm, whereas there is a clear degradation in the beam quality for the metallic samples, 10 nm Ag and 120 nm Au, seen in Figure \ref{fig:profile} (c) and (d) respectively. This is particularly evident in Au, for which a beam profile was entirely unattainable at L = 17.7 cm. Not presented are beam profiles for the remaining samples listed in Table \ref{tab:particles} (ZnO, TiO$_2$, and Fe$_3$O$_4$) which were only attained for L = 5.5 cm due to limited sample availability. The beam quality for the samples not presented are comparable to Ag at L = 5.5 cm (Figure \ref{fig:profile} (c)).

\begin{figure}[ht!]
\centering
\includegraphics[width=150mm]{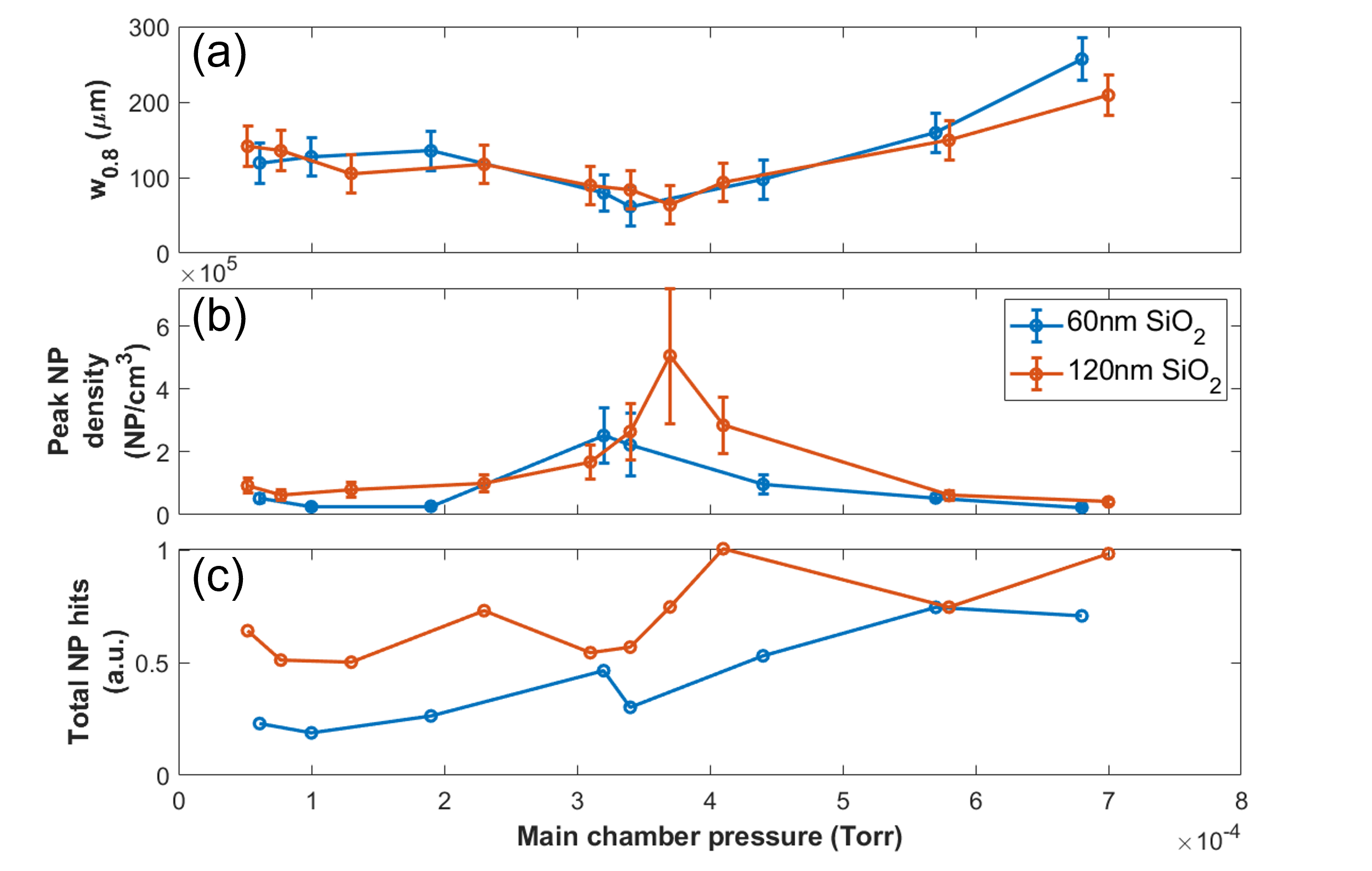}
\caption{60 nm and 120 nm SiO$_2$ NP beams: (a) w$_{0.8}$, (b) peak density, and (c) total NP hits as a function of gas flow into the aerodynamics lens as controlled by the pinch bolt (depicted in Figure \ref{fig:setup}) and monitored by the pressure in the interaction chamber. Error bars in (a) represent the step size of the motorized mirror, whereas those in (b) additionally account for the uncertainty in the focal spot size diameter.}
\label{fig:flow}
\end{figure}

To put a more specific value to the NP beam width for the type of scan presented in Figure \ref{fig:profile} we now define a beam width, w$_{0.8}$, such that w$_{0.8}$ is the range, centered about the center of mass of the NP hit profile, for which the integral of the profile in that range constitutes 80\% of all the NP hits for that scan. This definition is applied to perform a quantitative analysis of the NP beam width for 60 nm and 120 nm SiO$_2$ with respect to a variable gas flow into the aerodynamic lens and is presented in Figure \ref{fig:flow} (a). Here the gas flow is controlled by the pinch bolt (shown in Figure \ref{fig:setup}) and monitored by the pressure in the interaction chamber. It is important to note that w$_{0.8}$ is unaffected by reasonable choice of threshold.

\begin{figure}[t]
\centering
\includegraphics[width=\linewidth]{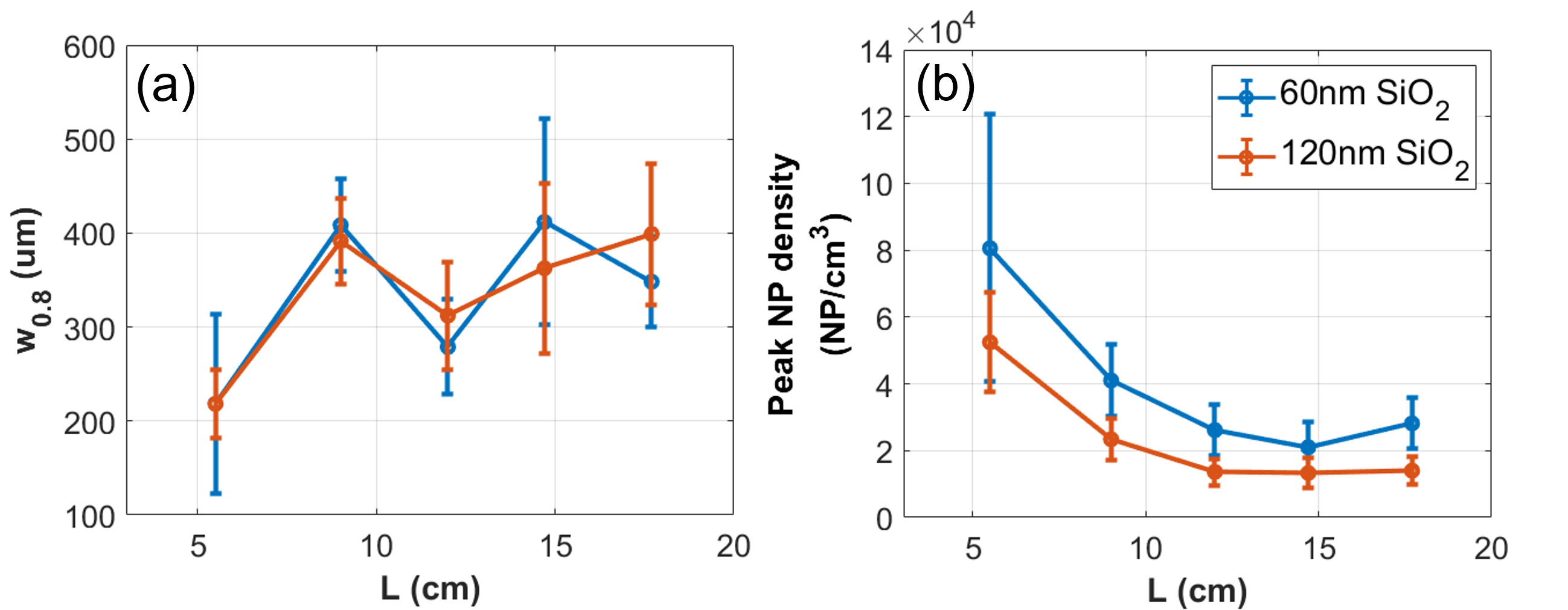}
\caption{NP beam characterization along its axis of propagation for 60 nm and 120 nm SiO$_2$ atomized at suspension concentrations of 5e13 and 6.3e12 NP/mL respectively. (a) presents w$_{0.8}$ for five L values between 5.5 cm and 18 cm and each 1D scan uses the same parameters as those used for the SiO$_2$ scans presented in Figure \ref{fig:concentration}. (b) shows the approximate in-vacuum NP volume density for these same five L values. This density is attained by modelling the interaction region as a cylinder of diameter equal to the 1/e$^2$ spot size of the laser focus and length equal to the w$_{0.8}$ of the NP beam. Error bars are computed by the same methods as in Figure \ref{fig:flow} (a) and (b) respectively.}
\label{fig:Lscan}
\end{figure}

The NP beam width analysis presented in Figure \ref{fig:flow} (a) is now utilized to approximate a particularly interesting value, the in-vacuum NP volume density (Figure \ref{fig:flow} (b)). By using the thresholding method we have attained the number of NP hits per laser shot and thus, by knowing the interaction volume, the NP volume density can be calculated. Once again, note that the limitations of the NP detection scheme imply that this approximation will be a lower bound to the actual NP density. In Figure \ref{fig:flow} (b) we present the peak NP density as a function of interaction chamber pressure. Specifically, we estimate the density for each pressure at the focal position corresponding to the maximum number of NP hits for the given 1D scan. The interaction region is approximated as a cylinder with diameter equal to the 1/e$^2$ spot size of the laser focus (45.8 ± 0.5 $\mu$m) and length equal to the w$_{0.8}$. Note that we use w$_{0.8}$ to be the length of the interaction region as the Rayleigh range of the laser focus is millimeters long, an order of magnitude larger than the effective width of the NP beam. Finally, Figure \ref{fig:flow} (c) shows the total number of NP hits for the scan performed at each pressure.
\\
The analysis described above culminates in a detailed picture regarding the behavior of the aerodynamic lens with respect to the input gas flow, where each of the three subfigures that comprise Figure \ref{fig:flow} provides key information. Figure \ref{fig:flow} (a) depicts a lens with a ‘preferred’ gas flow input, that is, the NP beam is focused mostly tightly at the gas flow rate corresponding to a pressure in the interaction chamber between 3e-4 and 4e-4 Torr for both 60 nm and 120 nm SiO$_2$. Consistent with this observation, Figure \ref{fig:flow} (b) then shows that the in-vacuum NP density is also maximized here. Finally, despite the NP beam being ‘optimized’ at these pressures, the total number of NPs comprising the beam is seen in Figure \ref{fig:flow} (c) as generally increasing with the gas flow throughput (as measured by chamber pressure). Though none of these observations are particularly surprising, Figure \ref{fig:flow}, as a whole, demonstrates two important points. First, the methods we have described thus far are self-consistent, validating the described SFI based technique as a reliable characterization method for an in-vacuum NP beam. Further, it demonstrates the technique as a powerful and practical tool when using an aerodynamic lens. Depending on the application an aerodynamic lens is being used for, a user may prefer a tight NP beam for a reliable target, whereas another application may require that the total number of NPs in-vacuum be maximized and a small target is of little consequence. The techniques presented here allow the user freedom to explore their relevant parameter space and tailor their NP beam to the specific application.
\\
Finally, a similar study is now applied to a variable L (Figure \ref{fig:Lscan}) which allows for a 3-dimensional characterization of the NP beam to be achieved. As seen in Figure \ref{fig:Lscan} (a) w$_{0.8}$ nearly doubles from L = 5.5 cm to L = 9 cm and then remains largely constant through the range allowed by the experimental apparatus. Figure \ref{fig:Lscan} (b) presents the complementary NP beam densities to the beam widths presented in Figure \ref{fig:Lscan} (a). It is immediately evident that the NP density decreases the further from the exit of the aerodynamic lens, dropping by a factor of $\sim$2 for both 60 nm and 120 nm SiO$_2$ from L = 5.5 cm to L = 18 cm. This is the case despite the NP beam width being largely constant with increasing L. Further, though the atomized NP samples vary by an order of magnitude for 60 nm and 120 nm SiO$_2$ (5e13 NP/mL and 6.3e12 NP/mL respectively), the maximum in-vacuum densities differ by less than a factor of 2.

Before giving our concluding remarks, we should mention that the SFI characterization technique is largely comparable to LIBD with the key difference being the detection of electrons (SFI) versus plasma photons (LIBD). Both detection schemes utilize a pulsed laser to determine the presence of a NP within the laser focus for each laser shot and are effective for smaller NP sizes ($\geq$ 10 nm). This gives both techniques access to the spatial dependence of the NP density with resolution equal to the laser focal volume and are subject to similar limitations. It is important to highlight some differences between SFI used here and LIBD as performed by Barreda et al.\cite{barreda2015situ}. First, we have made use of a fs laser which allows for focal peak intensities more than 1000 times larger than those produced by the ns laser used in [18]. This ensures that not only are the NPs in the focal region ionized, but they are ionized to an extent which allows electrons to be detected via CEM. Further, this extra peak power permits flexibility for an optical setup where the focusing lens is outside of the vacuum chamber. In our opinion, the biggest strength of SFI vs LIBD is that the latter relies on the emission of plasma photons from the NP which is a material dependent process and may not always be realizable, whereas ionization is possible with sufficient laser intensity, something very achievable with modest laser systems. This big advantage could make SFI a more general prospect compared to LIBD.

\section{Conclusion}

We have demonstrated the use of strong field ionization for the detection of isolated NPs in vacuum and the utility of this technique for 3-dimensional characterization of NP beams produced by an aerodynamic lens. By using a CEM in conjunction with a fs pulsed laser focused into the NP beam we determine if there is a NP present in the laser focus for each laser pulse by detecting its ejected electrons. With this information and with a known laser focal volume the NP density in the focus is determined. Scanning the laser focus through the NP beam allows for simple and effective mapping of the 3-dimensional NP density distribution. As discussed in the text, the thresholding technique undercounts the number of events in the beam and thus the calculated NP beam density represents a lower bound to the real NP density.
\\
Finally, we have applied this scheme to optimize the NP beam. The parameter to optimize in this instance is the aerosol flow into the aerodynamic lens. It is shown that the aerosol input influences both the collimation (width) of the NP beam and the number of NPs that comprise the beam, both of which effect the NP volume density and is crucial to many applications.

\section{Acknowledgments}

The authors would like to recognize the Air Force Office of Scientific Research (AFOSR) for financial support of this work under grant no. FA9550-21-1-0387. T.S. was partially funded by the U.S. Department of Energy (DOE) under grant no. DE-SC0019098.

\section{Author contributions statement}

M.D., T.S., N.G.H., J.A.P., and C.T. conceived the technique. M.D. and J.A.P. constructed the experimental apparatus.  M.D. and T.S. conducted the experiments. M.D. analysed the results and prepared the manuscript.  All authors contributed in discussions and reviewed the manuscript.

\section{Availability of data and materials}

The datasets used/analyzed in this manuscript are available from the corresponding author on reasonable request.

\section{Additional information}

The authors declare no competing financial interests.

\bibliographystyle{unsrt}  
\bibliography{references}

\end{document}